\documentclass[twocolumn,superscriptaddress,prb,showpacs,amsmath]{revtex4}
\usepackage{amsmath}
\usepackage{graphicx}
\usepackage{dcolumn}
\usepackage{bm}
\usepackage{color}

\begin{document}
\title{Topological quantum phase transition in the BEC-BCS crossover  phenomena}
\author{Mitsuhiro Arikawa}
\affiliation{Institute of Physics, University of Tsukuba 1-1-1 Tennodai, Tsukuba Ibaraki 305-8571, Japan}
\author{Isao Maruyama}
\affiliation{Graduate School of Engineering Science, Osaka University, 1-3 Machikaneyama-cho,
Toyonaka, Osaka 560-8531, Japan}
\author{Yasuhiro Hatsugai}
\affiliation{Institute of Physics, University of Tsukuba 1-1-1 Tennodai, Tsukuba Ibaraki 305-8571, Japan}

\date{\today}
\begin{abstract}
A crossover between the 
Bose Einstein condensation (BEC) and BCS superconducting state 
is described 
topologically 
in the chiral symmetric fermion system with 
attractive interaction.
Using a local $Z_2$ Berry phase,
we found a quantum phase transition 
between the BEC and BCS phases
without accompanying the bulk gap closing.
\end{abstract}
\maketitle

{\itshape Introduction}-
Recent progress in ultracold atomic Fermi gases,
the Feshbach resonance
which controls strength and sign of effective interaction,
 realizes 
the  Bose-Einstein Condensation (BEC)-BCS
 crossover\cite{regal,bartenstein,zwierlein}  
---  Bose condensation of bosonic molecules which are realized in 
real space pairing of fermions
and  the BCS superfluid with Cooper pairing in momentum space.
A long time ago, Leggett \cite{Leggett} proposed
 a trial wavefunction for the ground state  which describes not only 
weakly attractive case where formation of the Cooper pairs occurs,
but also the strong attractive case
where a dilute gas of molecules undergoes the BEC. 
Also substantial number of  theoretical works on the BEC-BCS crossover
have been performed\cite{Nozieres,Randeria,ohashi} after this pioneering work.  
By using a Ginzburg-Landau type argument based on the 
symmetry breaking and local order parameter,
these states are not clearly distinguished.
They are considered as a crossover, that is,
the two ground states are 
adiabatically connected even in the thermodynamic limit.

Both of the BCS and the BEC ground states are 
typical examples of gapped quantum states, 
which we call quantum liquids,
in the sense that there are no symmetry breaking phase 
transition between them. 
Still they are quite different and characteristic quantum states. As
discussed in this paper, there are clearly distinguished by using a
topological quantities. Therefore they belong to a new class of
matter as the {\it topological insulator} based on a novel classification
scheme (topological classification).

A class of quantum Hall states is a typical example of
topologically non trivial quantum liquids
where topological quantities play fundamental roles
for the 
characterization\cite{laughlin81,tknn82,wen89,hatsugai93}.
Also a time reversal invariant analogue of the quantum Hall states
are studied intensively\cite{kane,konig}.

As for the quantum liquids with non trivial topological structure,
we are proposing to use
topological quantities such as 
the Berry phases and the Chern numbers
using Berry 
connections\cite{hatsugai2004,maruyama2008,hirano2007a,hatsugai2010}. 

An important characteristic feature of the topological insulators
is an appearance of local degrees of freedom near boundaries
and impurities as generic edge 
states\cite{laughlin81,halperin82,hatsugai93,hatsugai93a}.
 Although the bulk is gapped and featureless, the edge states
characterize the topologically non trivial bulk. 
They are not independent but intimately related each other, 
which is known as 
 {\itshape bulk-edge correspondence}, where
topologically non-trivial bulk guarantees the existence of
localized modes and such low energy localized excitations characterize
the gapped bulk insulator\cite{hatsugai93} conversely.
The quantum Hall state of the graphene  
also belongs to this topological insulator
where the bulk-edge correspondence is important 
for the description
\cite{hatsugai2006a,arikawa2008,hatsugai2008rev}. 
The bulk-edge correspondence is also realized as the existence of 
the Kennedy triplet
for an open integer chain\cite{kennedy90}.
Such characteristic edge modes appear 
in the  valence bond solid (VBS) states\cite{katsura2007}
 and spin ladder with cyclic exchange 
interaction\cite{maruyama2008,arikawa2009}.
 
In this paper, we introduce a local $U(1)$ twist for the Hamiltonian
and define a Berry phase using its many body ground state,
which is quantized into $Z_2$ due to the chiral symmetry of 
the Hamiltonian
\cite{hatsugai2004,hatsugaiunp}. 
Although the bulk gap is adiabatically connected between the BEC-BCS
crossover, the gap of the twisted system may collapse at some value
between the BEC-BCS crossover. It is allowed since the
gap is not a thermodynamical property and can be collapsed by the 
local perturbation. 
It actually occurs in the BEC-BCS crossover. Then
the crossover of the bulk is distinguished by 
the {\it local} quantum phase transition.

\begin{figure}
\begin{center}
\includegraphics[width=1.9in]{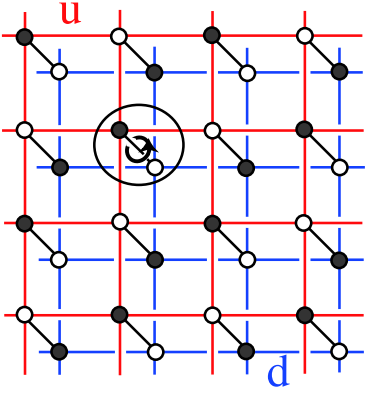}
\end{center}
\caption{Lattice structure of the effective mean field system for the two dimensional case.
We introduce a twist on the bond defined in Eq.(\ref{localdef}).
 \label{lattice}}
\end{figure}

{\itshape Model and $Z_2$ Berry phase }- 
Let us start from
the attractive Hubbard model at half-filling and discuss
a mean field Hamiltonian
\begin{eqnarray*}
{\mathcal H} & = & - t  \sum_{\sigma=\uparrow,\downarrow} \sum_{<{\mathbf i},{\mathbf j}>}  c^\dagger_{{\mathbf i},\sigma} c_{{\mathbf j},\sigma} -| U | \sum_{{\mathbf i}} n_{{\mathbf i},\uparrow} n_{{\mathbf i},\downarrow},
\end{eqnarray*}
where $c_{{\mathbf i},\sigma}$ is annihilation operator at site ${\mathbf i}$ with spin $\sigma$.
The number of sites is $N$.
The summation over $<{\mathbf i},{\mathbf j}>$ is restricted to the nearest neighbor pairs.
 The density with $\sigma$-spin at site ${\mathbf j}$
 is given by $n_{{\mathbf j},\sigma} = c^\dagger_{{\mathbf j},\sigma} c_{{\mathbf j},\sigma}$. 
We impose the periodic boundary condition.
We assume the lattice is bipartite in 1 to 3 dimensions.
To make the discussion clear, 
let us perform the following particle-hole transformation
in the mean field hamiltonian $H_{\rm MF}$, 
$
u_{\mathbf i}  = d_{{\mathbf i},\uparrow} = c_{{\mathbf i}, \uparrow}
$
and 
$
d_{\mathbf i}  =   d_{{\mathbf i},\downarrow} = c^\dagger_{{\mathbf i}, \downarrow}. 
$ 
(see Fig.\ref{lattice}), 
\begin{eqnarray}
{\mathcal H}_{\rm MF} & = &\sum_{<{\mathbf i},{\mathbf j}>}   -t  u^\dagger_{{\mathbf i}}  u_{{\mathbf j}}  + 
t d^\dagger_{{\mathbf i}}  d_{{\mathbf j}}  
 + \sum_{{\mathbf j}} \Delta \left(u^\dagger_{\mathbf j} d_{\mathbf j} + d^\dagger_{\mathbf j} u_{\mathbf j} \right).
\label{ham}
\end{eqnarray}
The order parameter of the superconductivity
$\Delta = -| U | \langle c^\dagger_{{\mathbf i},\uparrow} c^\dagger_{{\mathbf i},\downarrow} \rangle  = -|U| \langle u^\dagger_{\mathbf i} d_{\mathbf i} \rangle$, 
is chosen as a non-negative real number. 
In this paper, we do not perform the self consistent calculation and just assume
that the order parameter $\Delta$ is finite.
Here the expectation values $\langle {\mathcal O} \rangle$ is taken for the ground state.
The one-body eigenvalue problem of the $N$-site system is given by ${\mathcal H}_{\rm MF} | \phi^{(j)} \rangle  = E^{(j)}  | \phi^{(j)} \rangle$ ($E^{(1)} \le E^{(2)} \le \cdots \le E^{(2N)}$), 
where the  one-particle eigenstate is $ | \phi^{(j)} \rangle = (\phi^{(j)})^\dagger | 0 \rangle =\left [ 
\sum_{\mathbf m} \phi_{\mathbf m}^{(j,u)} u^\dagger_{\mathbf m} + \phi_{\mathbf m}^{(j,d)} 
d^\dagger_{\mathbf m} \right] | 0 \rangle$ 
with an orthogonal normalization  condition $\sum_{{\mathbf m},\alpha} 
\phi_{\mathbf m}^{(i,\alpha)}{}^* \phi_{\mathbf m}^{(j,\alpha)} =\delta_{ij}$.
Here the vacuum state $| 0 \rangle$ is defined as
 $u_{\mathbf j} | 0 \rangle = 0 = d_{\mathbf j} | 0 \rangle $ for any site ${\mathbf j}$. The $M$-particle eigenstate
 is constructed as $| \phi \rangle_M = \prod_{j=1}^M  (\phi^{(j)})^\dagger | 0 \rangle$ with the eigenvalue $\sum_{j=1}^M E^{(j)}$.  
At half-filling band, by Fourier transform it is easily found that the ground state $| \phi \rangle_N$ has a 
finite excitation gap energy $2\Delta$ for finite interaction $U$ for any bipartite system. 

\begin{figure}
\begin{center}
\includegraphics[width=3.5in]{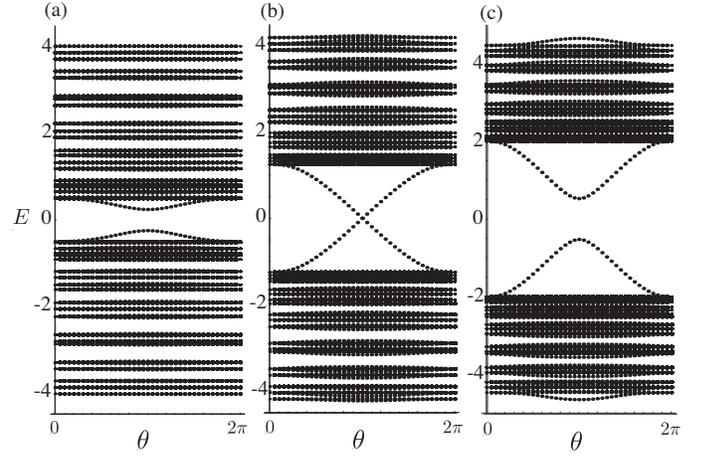}
\end{center}
\caption{Energy spectrum of ${\mathcal H}_{\rm MF}(\theta)$ for the two dimensional case ($N=16 \times 16)$: (a) $\Delta =0.5$. BCS phase. (b) $\Delta \simeq 1.250$. The level cross occurs at $(\theta, E) = (\pi,0)$.  (c) $\Delta =2$. BEC phase.
 \label{energyspectrum}}
 \end{figure}
 
 
The BCS and BEC states are not clearly distinguished 
and they are adiabatically connected as a crossover
 even in the 
thermodynamic limit. 
In order to define the Berry phase,  
 we modify the terms locally only at ${\mathbf j}={\mathbf a}$ in Eq.(\ref{ham}) as follows,
 \begin{eqnarray*}
 \Delta \left(u^\dagger_{{\mathbf a}} d_{{\mathbf a}} + d^\dagger_{{\mathbf a}} u_{{\mathbf a}} \right) 
 & \rightarrow & 
 \Delta \left(e^{i \theta} u^\dagger_{{\mathbf a}} d_{{\mathbf a}} +
 e^{-i \theta} d^\dagger_{{\mathbf a}} u_{{\mathbf a}} \right).
 \label{localdef}
 \end{eqnarray*} 
 Hereafter we take $t$ as unity.
 Figure~\ref{energyspectrum} presents the $\theta$-dependence of the energy spectrum of Eq.(\ref{mffluxhamiltonian}) 
for the square lattice.
The energy gap $\delta E = E^{(N+1)}-E^{(N)}$ minimizes at $\theta=\pi$.
In Fig.\ref{energyspectrum} (b), we can see the level-cross occurs at $(\theta,E)=(\pi,0)$ around $\Delta = \Delta_c \simeq 
1.250 $
for two dimensional case. 
As we will see later, $\Delta=\Delta_c$ is the {\itshape critical point} 
between the BEC and BCS phase of the BEC-BCS crossover, although the gap of the translational invariant 
system ($\theta=0$) is always open.  

 Due to this local perturbation, the spectrum changes, which can be regarded as the edge mode.
It characterizes the
gapped bulk feature as a realization of the bulk-edge correspondence.
Under the above deformation, the half-filled eigenstate $| \phi \rangle_N$ depends on the $\theta$, which denotes $| \phi (\theta) \rangle_N$. 
The Berry phase $\gamma$ is then 
defined as
\begin{eqnarray}
i \gamma = \int_0^{2\pi}
d\theta \, 
 {}_N\! \langle \phi (\theta) | \partial_\theta \phi (\theta) \rangle_N.
 \label{defBerry}
\end{eqnarray}

 \noindent 
The transformation $(u_{\circ},d_{\bullet}) \rightarrow (- d_{\bullet}, u_{\circ})$ (see Fig.~\ref{lattice}) for one sublattice
reduces the Hamiltonian in the  following bilinear form:
\begin{eqnarray}
{\mathcal H}_{\rm MF}(\theta) 
& = & \sum_{{\mathbf i},{\mathbf j}} u^{\dagger}_{\mathbf i}
  [D_N(\theta)]_{{\mathbf i}{\mathbf j}} d_{\mathbf j} + {\rm H.c.},
\label{mffluxhamiltonian}
\end{eqnarray}
This Hamiltonian (\ref{mffluxhamiltonian}) has a chiral symmetry,
$\{H_{\rm MF}, ^\exists\gamma\}=0$, $\gamma^2=1$.
Unless the determinant of the matrix $D_N(\theta)$ vanishes, the half-filling ground state of the Hamiltonian 
(\ref{mffluxhamiltonian}) has a finite gap. 
Therefore, the Berry phase $\gamma$
is quantized as $0$ or $\pi$ modulo $2\pi$ ($Z_2$ Berry phase)
\cite{hatsugai2004}.
Using this $Z_2$ Berry phase, 
we identify whether the half-filled ground state is 
a  BEC or BCS state. 
The Berry phase (\ref{defBerry}) for the 
half filled ground state of the chiral symmetric  Hamiltonian
 (\ref{mffluxhamiltonian})
can be obtained as 
following expression\cite{hatsugaiunp}: 
\begin{eqnarray}
\gamma & = & \int_0^{2\pi} d \theta \ {\rm Im} \log \det D_N(\theta). 
\label{hatsugaieq}
\end{eqnarray}
By definition, we can show that
the $Z_2$ Berry phase remains invariant through an adiabatic deformation
until the level cross occurs.
That is, the Berry phase is topologically protected.
In this paper
we evaluate the $Z_2$ Berry phase intuitively using
an adiabatic continuation without direct calculation of Eq.(\ref{hatsugaieq}).

\begin{figure}
\begin{center}
\includegraphics[width=3.2in]{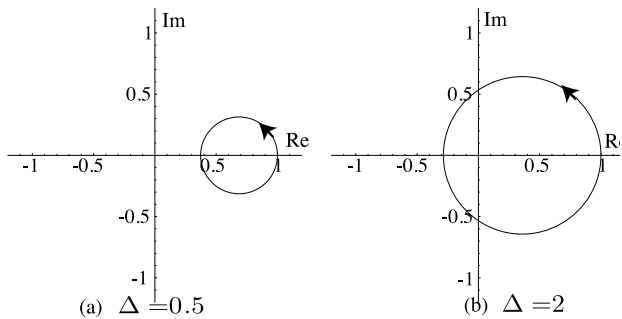}
\end{center}
\caption{The $\theta$-dependence of $\det D_N(\theta)/\det D_N(0)$
 in the complex plane for $8 \times 8$ square lattice case.
(a) $\Delta=0.5$ (b) $\Delta=2$. 
 \label{thetadep}}
 \end{figure}

{\itshape Adiabatic deformation}-
We consider the two types of the adiabatic deformation to understand the BCS and BEC states explicitly.
Since we introduce the flux $\theta$ 
on one local bond, it turns out that $\det D_N(\theta)$ has a form $ A e^{i \theta} + B$ ($A$ and $B$ are
real and independent of $\theta$) by the  Laplace expansion.  
For a finite attraction $U$ (i.e., finite $\Delta$), only at $\theta=\pi$, $\det D_N(\theta)$ can become zero 
where the level-cross occurs at zero energy.
During the change of the variable $\theta$ from $0$ to $2\pi$, the determinant $\det D_N(\theta)$ draws a closed curve in the complex plane as shown in Fig.\ref{thetadep}.  
If the $\det D_N(\theta)$ winds around the origin of the complex plane $m$
times when $\theta$ varies from $0$ to $2\pi$, 
the Berry phase $\gamma$ is given by $\gamma = \pi m$ modulo $2\pi$.  
When the coupling strength satisfies that $0 < \Delta < \Delta_c$, 
the closed curve of $\det D_N(\theta)$ does not enclose the origin (see Fig.\ref{thetadep}(a)). 
From Eq.(\ref{hatsugaieq}), we obtain the Berry phase $\gamma$ as $\gamma=0$.
When $\Delta > \Delta_c$, 
 the closed curve $\det D_N(\theta)$ winds origin once (see Fig.\ref{thetadep} (b)). Then, we obtain $\gamma=\pi$.

 \begin{figure}
\begin{center}
\includegraphics[width=3.2in]{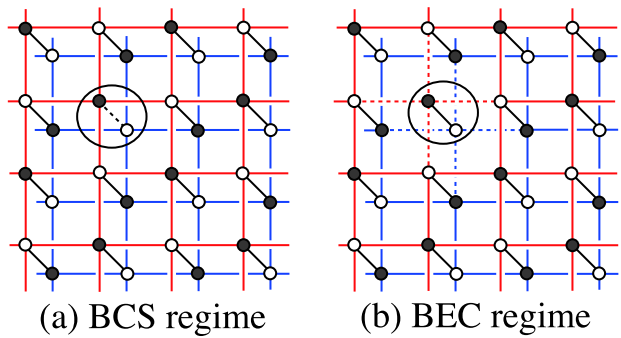}
\end{center}
\caption{
Two types of the adiabatic deformations. One deformation is the decrease of the local bond (type (i)) (a).
The other is the decrease of the hopping connecting the deformed  local bond being zero (type (ii)) (b).
 \label{deformation}}
\end{figure}
 
 \begin{figure}
\begin{center}
\includegraphics[width=2.3in]{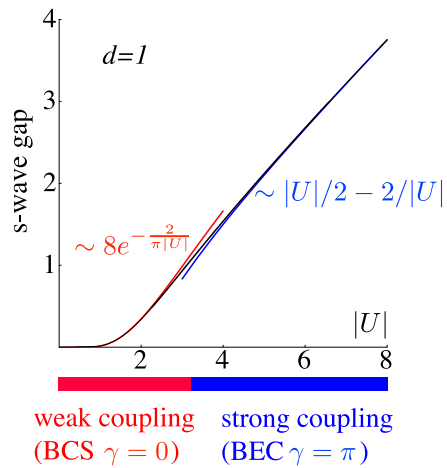}
\end{center}
\caption{Relation between the $s$-wave gap $\Delta$ and the attractive interaction strength $|U|$
in one dimension.
 In weak coupling (BCS) regime, $\Delta$ behaves $\Delta \sim 8 \exp -2/(\pi |U|)$, where the Berry phase is $0$. 
 In strong coupling (BEC) regime, $\Delta \sim |U|/2 - 2/|U|$, where the Berry phase is $\pi$.  
 \label{gapandberry}}
\end{figure}

 {\itshape Type (i) }:
 As seen in Fig.~\ref{thetadep}(a), for small $\Delta (=0.5$), $\det D_N(\theta)$ draws the circle without surrounding 
the origin in the complex plane. 
Then we adiabatically decrease the magnitude of the twist interlayer hopping $\Delta$ at ${\mathbf j}={\mathbf a}$ to zero (see Fig.~\ref{deformation} (a)).
During this adiabatic deformation,
the circles in the complex plane shrink to the one point in a concentric fashion without touching the origin
as the interlayer hopping $\Delta$ on the twist bond goes to zero (i.e., during the adiabatic deformation 
$A \rightarrow 0$ and $B$ remains same
 when we write down $\det D_N(\theta)$ as $\det D_N(\theta) = A e^{i \theta}+ B$).  
During this deformation, the excitation gap of the half-filling ground state remains finite.
After the deformation, the dependence of $\theta$ disappears in Eq.(\ref{mffluxhamiltonian}). 
Due to the invariance of the $Z_2$ Berry phase under 
the adiabatic deformation, the Berry phase $\gamma$ should be zero for small $\Delta$, where
 the binding of the pairing is in momentum space (BCS state).   
  In contrast, during the deformation, the circle touches the origin in the complex plane
 once for large $\Delta(=2.0$).
 For large $\Delta$ the level-cross occurs at zero energy for large $\Delta$.    
 At the level crossing point during the deformation, the Berry phase is not protected any more.  
 To understand the Berry phase for large $\Delta$ we consider another type deformation.
 
\begin{table}
\begin{center}
\caption{Critical values $\Delta_c$ for dimensions $d=1,2$ and $3$.\label{tablecritical}}
\begin{tabular}{| c | c | }
\hline
Dimension $d$ & $\Delta_c$ \\
\hline
$d=1$(linear chain) & $2/\sqrt{3} \simeq 1.1547$ \\
$d=2$(square lattice) & $1.250$ \\
$d=3$(cubic lattice) & $1.588$ \\
\hline
\end{tabular}
\end{center}
\end{table} 
 {\itshape Type (ii)}: We decrease the hopping $t$ connecting the site ${\mathbf a}$ into zero.
 Figure \ref{thetadep}(b) shows the  $\det D_N(\theta)$
drawing the circle with surrounding the origin once in the complex plane. 
 In this deformation, the circle is parallel shifted so that the center of the circle becomes the origin of the complex plane  
 i.e., $A$ remains same and $B \rightarrow 0$ for $\det D_N(\theta) = A e^{i \theta}+ B$.
During this deformation the level-cross does not occur at zero energy for large $\Delta$.
After the deformation, the twist interlayer hopping $\Delta$ at ${\mathbf j}={\mathbf a}$ becomes isolated. 
Then, it is easy to show the isolated interlayer bond gives the $\pi$ Berry phase.
Thus, the two electrons form a dimer bound state (BEC state) for large $\Delta$. 
When $\Delta$ is small, the Berry phase is zero as discussed above.
During the deformation of type (ii), the circles touch the origin in the origin, that is, the level-cross occurs at zero energy.
From numerical calculation we find that there exists only 
 one critical point $\Delta = \Delta_c$, where the Berry phase is $0$ ($\pi$) for 
 $\Delta < \Delta_c$ ($\Delta > \Delta_c$). 
 In a similar manner, the critical value $\Delta_c$ for the other dimensions $d$ can be calculated (see Table \ref{tablecritical}). 
 It is obtained numerically as $\Delta_c \simeq 1.588$ for the cubic lattice. 

In one dimensional case, the expression of $\det D_N(\theta)$ can be obtained analytically as
\begin{eqnarray*}
{\det D_N(\theta)} =
i \Delta e^{i \theta}  U_{N-1}\left( \frac{i \Delta}{2} \right) -2 U_{N-2}\left( \frac{i \Delta}{2} \right) -2.
\end{eqnarray*}
Here the $U_n(x)$ is the Chebyshev polynomial of the second kind\cite{nash86}, 
defined as $U_n(\cos x)={\sin[(n+1) x]}/{\sin  x}$.
Parametrizing $\Delta$ by $\Delta= 2 \sinh \alpha$,
the asymptotic form of the $\det D_N(\theta)$ for large $N$ can be expressed as
\begin{eqnarray*}
\det D_N(\theta) 
& \simeq &\left( \frac{i \Delta e^\alpha}{2\sinh \alpha}\right)^N \frac{e^{i \theta}\sinh \alpha + e^{-\alpha}}{\cosh \alpha}.
\label{asympdet}
\end{eqnarray*}
We have a real root on $\alpha$ for the equation $\det D_N(\theta) =0$, i.e.,  $e^{i \theta}\sinh \alpha + e^{-\alpha}=0$ only when $\theta=\pi$. The solution is obtained as $\alpha=(\log 3)/2$, i.e., 
$\Delta_c=  2/\sqrt{3} \simeq 1.1547$, although the bulk gap $(\theta=0)$ remains finite. 
 
For one dimensional system, 
the open boundary chain of Hubbard model are analyzed 
\cite{frahm,deguchi},
which 
might be relevant to characterize the bulk features.

{\itshape Summary and discussions}-
We show the quantum phase transition in the mean-field attractive Hubbard model at half-filling.
The $Z_2$ Berry phase distinguishes the BEC-BCS crossover as a local quantum 
phase transition, that is, the phases are separated by closing of the energy 
gap under the local twist, although the gap of the translational invariant 
system is always open. 
In the weak attractive interaction (BCS) case, 
the Berry phase is $0$ in one hand. 
On the other hand, 
in the strong attraction (BEC) case, the Berry phase is $\pi$ (see Fig.~\ref{gapandberry}). 	
That is, it characterizes whether the paired electron is 
itinerant or localized. This comes from the bulk-edge correspondence 
in the BEC-BCS crossover.
Physically 
the itinerant Cooper pairs in the BCS phase
 are not affected by the local $U(1)$ twist
although the spatially localized bosons formed by the real space 
is
crucially affected by the twist which results in the non trivial $\pi$ Berry 
phase. In this sense,  our topological characterization reflects
whether the size of the pairing is macroscopic or of the order unity.


{\itshape Acknowledgements}-
The work was supported in part by Grants-in-Aid for Scientific 
Research, Grant No.20654034 from JSPS and No.20029004 (Physics 
of New Quantum Phases in Super-clean Materials) and 
No.20046002 (Novel States of Matter Induced by Frustration) 
on Priority Areas from MEXT (Japan).
Some numerical calculations were carried out on Altix3700BX2 at YITP in Kyoto University and 
the facilities of the Supercomputer Center, Institute for Solid State Physics, University of Tokyo.


\begin{thebibliography}{99}

\bibitem{regal}C.A.~Regal, M.~Greiner, and D.S.~Jin, Phys. Rev. Lett. {\bf 92}, 040403 (2004).

\bibitem{bartenstein}M.~Bartenstein, A.~Altmeyer, S.~Riedl,  S. Jochim, C. Chin, J. H. Denschlag, and R. Grimm,
 Phys. Rev. Lett. {\bf 92}, 120401 (2004).
 
 \bibitem{zwierlein}M. W. Zwierlein, C. A. Stan, C. H. Schunck, S. M. F. Raupach, A. J. Kerman, and W. Ketterle,
  Phys. Rev. Lett. {\bf 92}, 120403 (2004).

\bibitem{Leggett}A. J. Leggett, J. Phys. (Paris) {\bf 41}, C7 (1980).

\bibitem{Nozieres}P. Nozieres and S. Schmitt-Rink, J. Low Temp. Phys. {\bf 59}, 195 (1985). 
\bibitem{Randeria}M. Randeria, in {\itshape Bose-Einstein condensation},
A. Griffin, D. W. Snoke, and S. Stringari (eds.), Cambridge University Press, Cambridge (1995). 

\bibitem{ohashi}Y. Ohashi and A. Griffin, Phys. Rev. Lett. {\bf 89}, 130402 (2002).


\bibitem{laughlin81}R.B.~Laughlin, \prb {\bf 23}, 5632 (1981). 



\bibitem{tknn82}D.J.~Thouless, M.~Kohmoto, M.P.~Nightingale and M.~den Nijs, \prl {\bf 49}, 405 (1982). 

\bibitem{wen89} X. G.~Wen, \prb {\bf 40}, 7387 (1989). 

\bibitem{hatsugai93} Y.~Hatsugai, \prl {\bf 71}, 3697 (1993). 

\bibitem{kane}C. L. Kane and E. J. Mele, Phys. Rev. Lett. {\bf 95}, 146802 
(2005); {\bf 95}, 226801 (2005).

\bibitem{konig} M. K\"{o}nig, S. Wiedmann, C. Br\"{u}ne, A. Roth, H. Buhmann, L. W. Molenkamp, X.-L. Qi, S.-C. Zhang, Science {\bf 318}, 766 (2007).

\bibitem{hirano2007a}
T.~Hirano, H.~Katsura, and Y.~Hatsugai, \prb {\bf 77}, 094431 (2008); \prb {\bf 78}, 054431 (2008). 

\bibitem{maruyama2008}I.~Maruyama, T.~Hirano, and Y.~Hatsugai, 
\prb {\bf 79}, 115107 (2009).


\bibitem{hatsugai2004} Y.~Hatsugai, J. Phys. Soc. Jpn. {\bf 75}, 123601 (2006).
\bibitem{hatsugai2010} Y.~Hatsugai, New J. Phycs.,  to appear (2010).

\bibitem{halperin82}B. I.~Halperin, \prb {\bf 25}, 2185 (1982). 

\bibitem{hatsugai93a} Y.~Hatsugai, \prb {\bf 48}, 11851 (1993). 

\bibitem{hatsugai2006a}Y.~Hatsugai, T.~Fukui, and H.~Aoki, \prb {\bf 74}, 
205414 (2006). 

\bibitem{arikawa2008}M.~Arikawa, Y.~Hatsugai, and H. Aoki, \prb {\bf 78}, 205401 (2008).

\bibitem{hatsugai2008rev}Y. Hatsugai, Solid State Comm. {\bf 79} 205107 (2009). 

\bibitem{kennedy90}T.~Kennedy, J.Phys: Condens. Matter {\bf 2}, 5737 (1990).

\bibitem{katsura2007}H.~Katsura, T.~Hirano, and Y.~Hatsugai, \prb {\bf 76}, 012401 (2007).  


\bibitem{arikawa2009}M.~Arikawa, S.~Tanaya, I.~Maruyama, Y.~Hatsugai
Phys. Rev. B {\bf 79}  205107(2009).

\bibitem{hatsugaiunp}Y.~Hatsugai and I.~Maruyama, in preparation. 


\bibitem{nash86}T.S. Chihara, {\itshape An Introduction to Orthogonal Polynomials}, Gordon and Breach, New York (1978). 



\bibitem{frahm}G.~Bed\"{u}rftig and H.~Frahm, J. Phys. A: Math. Gen. {\bf 30}, 4139 (1997).

\bibitem{deguchi} T.~Deguchi, R.~Yue, and K.~Kusakabe, J. Phys. A: Math. Gen. {\bf 31}, 7315 (1998).

\end{thebibliography}
\end{document}